\documentclass[11pt]{article}

\usepackage{graphicx,amsfonts,latexsym}

\usepackage{eurosym}

\usepackage[english]{babel}
\usepackage{amssymb}
\usepackage{amsmath}
\usepackage{latexsym}
\usepackage{graphicx}
\usepackage{a4wide}

\newcommand{\qed}{\hfill $\square$}

\renewcommand{\d}{\mbox {\rm d}}

\newcommand{\Prob}{\mathbb{P}}

\newcommand{\beq}{\begin{equation}}
\newcommand{\eeq}{\end{equation}}
\newcommand{\bea}{\begin{eqnarray}}
\newcommand{\eea}{\end{eqnarray}}
\newcommand{\beas}{\begin{eqnarray*}}
\newcommand{\eeas}{\end{eqnarray*}}
\newcommand{\ba}{\begin{array}}
\newcommand{\ea}{\end{array}}
\newcommand{\bit}{\begin{itemize}}
\newcommand{\eit}{\end{itemize}}
\newcommand{\ben}{\begin{enumerate}}
\newcommand{\een}{\end{enumerate}}

\newcommand{\E}[1]{ { {\mbox{E}}  \left\{{#1}\right\}} }

\newtheorem{theorem}{Theorem}

\newtheorem{remark}{Remark}

\begin{document}
\author{Giuseppe Carlo Calafiore\thanks{Giuseppe C. Calafiore is  professor of Automatic Control at Dipartimento di Automatica e Informatica,
Politecnico di Torino, Italy.
Tel.: +39-011-090.7071; Fax:
+39-011-090.7099. E-mail: {\tt giuseppe.calafiore@polito.it}
}
}
\title{The Dating Problem and  Optimal Ordering \\ of  Sequential Opportunities}

%
\date{}
\maketitle

\begin{abstract}
In this paper, we discuss a stochastic decision problem of optimally selecting the order in which to try $n$ opportunities
 that may yield an uncertain reward in the future. The motivation came out from pure curiosity, after an informal
 conversation about  what could be the best way to date friends. The problem structure turned out to be suitable also for other situations, such as the problem of optimally selecting the order of submission of a paper to journals.
 Despite the seemingly combinatorial nature of the problem, we show that optimal-tradeoff solutions can be found
 by simply ordering a sequence of $n$ real numbers. 
 \vspace{.1cm}

\end{abstract}

\section{Introduction}
This paper deals with a sequential stochastic decision problem involving the optimal scheduling of a set of $n$ available ``opportunities,'' whose outcome is uncertain and delayed in time. The problem setup is best introduced via a specific example that we call the ``Dating Problem,'' which is exposed next.

\paragraph{The Dating Problem.} 
A version of this problem arose during a freewheeling conversation with my friend and colleague Laurent El Ghaoui, one night of September 2015 in Berkeley. A politically corrected narrative is as follows.
  A person (hereafter, the ``player'') has the contact numbers of $n$ friends.
At some time of the day $\tau_0=0$ the player can select a first friend $i\in\{1,\ldots,n\}$ 
 and send a message inviting her out for dinner.
Each friend has a ranking (i.e., value -- the larger the better) $r_i\geq 0$ in the player's opinion  or, equivalently, $r_i$ is the reward the the player gets if the 
$i$-th friend accepts the invitation. Also, each friend has a probability $p_i\in (0,1]$ of accepting the invitation, and  is characterized by the (random) time $\tau_i\geq 0$ she takes
for responding (positively or negatively) to the invitation. 
Once an invitation message is sent, the player enters a lock-in period in which she cannot send out other invitations, since she cannot incur the risk (and the corresponding embarrassment) of receiving multiple positive answers.
The lock-in period ends when the friend responds.
If the friend $i$ invited at $\tau_0$ responds at time $\tau_1$ then, in the case of a positive answer the player gets reward $r_i$ and the game ends, and in the case of a negative answer she can send out at time $\tau_1$ another invitation to a newly selected friend  $j\neq i$, and wait for her response. The process goes on this way until either {\em (a)} the player receives a positive answer, or
 {\em (b)} she receives negative answers from all the $n$ friends; we let $T$ denote the (random) time at which the game ends.
 The eventual reward $R$ that the player gets is equal to the ranking 
 of the friend that accepted the invitation, in case  {\em (a)}, or it is equal to zero in case {\em (b)}.
 Clearly,  $R$ is a discrete random variable with support $\{r_1,\ldots,r_n,0\}$, and both $R$ and $T$
 depend on the specific invitation sequence that the player selects. 
 The stochastic  decision problem the player faces is to determine an invitation sequence that optimizes
 a tradeoff between the expected reward (the higher the better) and the expected time at which the reward is obtained
 (the sooner the better).

\vspace{.5cm} 
The mathematical structure behind the  Dating Problem can actually  be applied to several other interesting situations. Two of them are outlined next, further discussion is delayed to Section~\ref{sec:discussion}.

\paragraph{The Journal Submission Problem.}
An author has a scientific paper ready for submission, and there are $n$ journals that may be suitable for submitting the paper to. Each journal has a reputation $r_i$, which may be given, for instance, by the impact factor of the journal. Also, from each journal's web page the author may find information about the average acceptance rate $p_i$
of submitted papers to the $i$-th journal, and on the average number of weeks that are needed  for receiving an editorial decision. Since parallel submissions are forbidden, the author should decide to which journal to submit her paper first. Then, if the firstly selected journal refuses the paper,  she may try with a second journal, and so on until the paper gets eventually accepted by some journal, or refused by all. Again, the problem is to optimally decide the submission sequence 
so to achieve an optimal tradeoff between expected reward and expected time to publication.

\paragraph{The House Bidding Problem.}
A family wants to buy a new house. There are $n$ interesting houses available in the real estate market, each of which has a ranking $r_i$ in the family's opinion.  For each of the $n$ houses it is possible to send 
a binding offer to the real estate agency dealing with that property. The processing of each offer takes a random time $\tau_i$, and the estimated probability of a positive outcome for the $i$-th house is $p_i$.
The family must decide on which house to bid first.


\subsection{Disclaimer}
We wanted to preserve in the manuscript the
spontaneous nature of the creative process: the dating problem was born as  
{\em divertissement}, and the solution presented here was worked out 
by proceeding directly, for the fun of it, without prior examination of related literature. 
After having obtained the main result, and at the time of putting things in clean form,
however, we did search the literature for related contributions and, not surprisingly, we found some. These are discussed briefly in the conclusion section.
 We  make no claim that the structure underlying the dating problem 
 is fully novel, nor that its solution cannot be possibly derived as a sub-case of some published result; only, that was not the route we took, and the specific setup and tradeoff optimization scheme we propose here do seem novel, to the best of the author's knowledge.

\section{Problem formulation and notation}
We next introduce a general framework in which all problems similar to the ones exemplified above may be cast.
We refer generically to friends, journals,  houses, etc. as ``opportunities.''
Each opportunity $O_i$, $i=1,\ldots,n$, is described
by a triple $O_i=(r_i,p_i,\tau_i)$,
where  $r_i\geq 0$ is the reward the the player may get
from the $i$-th opportunity,
 $p_i\in (0,1]$ is the probability 
 with which the opportunity actually returns its reward $r_i$ to the player (i.e., the positive outcome probability), 
 and $\tau_i\geq 0$ is the 
 random time at which the opportunity outcome is revealed (i.e., the opportunity's response time).
 
Once an opportunity is tried, the player enters a lock-in period in which she cannot try other opportunities (exclusive opportunities).
The lock-in period ends when the opportunity outcome is revealed. In the case of a positive outcome from the $i$-th opportunity, the player gets the corresponding reward $r_i$, and the game ends; in the case of a negative outcome, the player tries another opportunity, and so on until either a reward is obtained, or all opportunities have been tried with negative outcomes, in which case the final reward is zero.

We consider  the  response times $\tau_i$,  $i=1,\ldots,n$, to be independent. We
denote with $F_i(t)\doteq  \Prob\{\tau_i \leq t\}$  the cumulative distribution function (CDF)
 of  $\tau_i$,  and  we let $\theta_i$ denote the mean response time.
 We denote by $\sigma$ a permutation of the first $n$ integers $\{1,\ldots,n\}$
 representing the scheduling order of the opportunities, so that $\sigma(1)$ is the index of the opportunity that is tried first, $\sigma(2)$ the one tried second, etc.
We initially suppose that the scheduling sequence $\sigma$ is fixed, and that the game is started at time $\tau_0=0$ by trying the first opportunity $\sigma(1)$, and waiting for its outcome.

\begin{figure}[htb]
\centerline{\scalebox{1.6}{\includegraphics{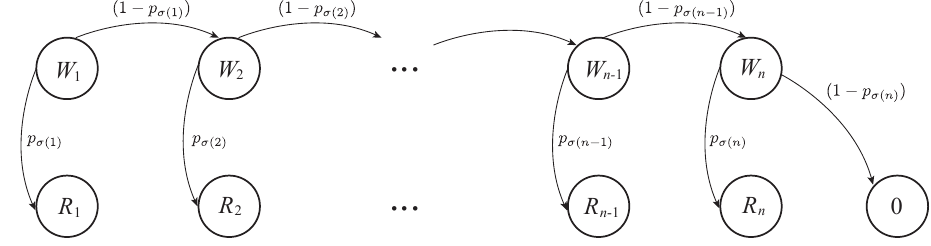}}}
\caption{State transition graph for the sequential opportunity game.}
\label{fig:states}
\end{figure}

The stochastic process representing the game is  denoted by $x(t)$,
and this process has a finite number of possible states, as depicted in Figure~\ref{fig:states}. The state $W_i$ represents the situation when the player is waiting for the outcome of opportunity $O_{\sigma(i)}$ (i.e., the $i$-th opportunity in the given scheduling sequence), and the state $R_i$
represents the situation when $O_{\sigma(i)}$ returns a positive outcome, hence the player  receives
reward $r_{\sigma(i)}$ and the game ends. State ``0'' represents the situation when the game ends with zero reward, which happens when all the opportunities returned a negative outcome.

\section{The reward and finish time distributions}
Consider the probability with which the process  $x(t)$ enters one of the absorbing states
$R_i$, $i=1,\ldots,n$, or the ``0'' state, in which the final reward is zero. We let
$\pi_0(t) \doteq \Prob\{x(t) = \mbox{``0''}\}$, and
\[
\pi_i(t) \doteq \Prob\{x(t) = R_i\},\quad i=1,\ldots,n.
\]
Further, we define the random variables
\[
s_k \doteq \sum_{i=1}^k \tau_{\sigma(i)},\quad k=1,\ldots,n,
\]
that describe the sum of the response times of the first $k$ opportunities in the scheduling sequence $\sigma(1),\ldots\sigma(n)$.
We denote by $\Upsilon_k(t) \doteq \Prob\{s_k \leq t\}$ the cumulative probability distribution function of
$s_k$. 
We observe that the event $\{x(t) = R_1\}$ happens when the first opportunity $O_{\sigma(1)}$ is tried,
a response is obtained at some time instant $\tau_{\sigma(1)} \leq t$, and this response is positive.
The probability of the event is thus $p_{\sigma(1)} \Prob\{s_1 \leq t\}$, that is
\[
\pi_1(t) = p_{\sigma(1)} \Prob\{s_1 \leq t\}= p_{\sigma(1)} \Upsilon_1(t).
\]
The event $\{x(t) = R_2\}$ happens instead when the first opportunity $O_{\sigma(1)}$ is tried,
a response is obtained at some time instant $\tau_{\sigma(1)} \leq t$, and this response is negative;
then the second opportunity $O_{\sigma(2)}$ is tried,
a response is obtained at some time instant $\tau_{\sigma(1)}+\tau_{\sigma(2)} \leq t$, and this response is positive.
The joint probability of these events is thus $(1-p_{\sigma(1)})p_{\sigma(2)} \Prob\{s_2 \leq t\}$, that is
\[
\pi_2(t) = (1-p_{\sigma(1)})p_{\sigma(2)} \Upsilon_2(t).
\]
Similarly, the probability pf the event $\{x(t) = R_3\}$ is 
\[
\pi_3(t) = (1-p_{\sigma(1)})(1-p_{\sigma(2)}) p_{\sigma(3)} \Upsilon_3(t),
\]
and proceeding along the same line one obtains that, for  $i=1,\ldots,n$,
\beq
\pi_i(t) = c_i \Upsilon_i(t),\quad \mbox{where } c_i\doteq p_{\sigma(i)}\cdot\prod_{j=1}^{i-1}(1-p_{\sigma(j)}),
\label{eq:pii}
\eeq
and
$
\pi_{0}(t) = \Prob\{x(t) = \mbox{``0''}\} =  c_{n+1} \Upsilon_{n}(t)
$,
where
$
c_{n+1}\doteq  \prod_{j=1}^{n}(1-p_{\sigma(j)})
$,
and it can readily verified that
$
\sum_{i=1}^{n+1} c_i = 1
$.
Note that $c_i$, $i=1,\ldots,n$, represents the probability of having the first success exactly at the $i$-th trial
in a sequence of independent Bernoulli trials having success probabilities $p_{\sigma(1)},\ldots,p_{\sigma(i)}$
(incidentally, the sum of Bernoulli trials with possibly unequal success probabilities give rise to a so-called Poisson's binomial distribution, see, e.g., \cite{Wang:93}).
Given a scheduling sequence $\sigma$, vector $\pi(t) = (\pi_1(t),\ldots,\pi_n(t),z(t))$
represents the probability mass distribution over the rewards $(r_{\sigma(1)},\ldots,r_{\sigma(n)},0)$,
where $z(t) \doteq 1- \sum_{i=1}^n \pi_i(t)$ is the probability of having zero reward at time $t$ (notice that this is the probability of being in state ``0,'' plus the sum of the probabilities of all the waiting states $W_1,\ldots,W_n$).
The expected reward at time $t$ is therefore expressed by
\beas
\E{R(t)} &=& \sum_{i=1}^n r_{\sigma(i)} \pi_i(t) = 
\sum_{i=1}^n r_{\sigma(i)} c_i \Upsilon_i(t)  
=\sum_{i=1}^n\Upsilon_i(t)   r_{\sigma(i)} p_{\sigma(i)}\cdot\prod_{j=1}^{i-1}(1-p_{\sigma(j)})  .
\label{eq:exp_reward}
\eeas
We observe that the expected reward is a linear combination of the CDFs $\Upsilon_i(t)$, with positive coefficients.
Since  $\Upsilon_i(t)$ is monotone nondecreasing and $\Upsilon_i(t)|_{t\to\infty}=1$, we conclude that $\E{R(t)}$ is also monotone nondecreasing,
and its maximum value is achieved in the limit for $t\to\infty$. We thus define the eventual expected reward $\bar R$ as
\beq
\bar R = \bar R(\sigma) \doteq \lim_{t\to\infty} \E{R(t)} = \sum_{i=1}^n r_{\sigma(i)} c_i = 
\sum_{i=1}^n  r_{\sigma(i)} p_{\sigma(i)}\cdot\prod_{j=1}^{i-1}(1-p_{\sigma(j)})  .
\label{eq:ssexpreward}
\eeq
Next, we define the random variable $T$ as the time at which the game ends, which coincides with the time at which the process $x(t)$ enters one of the absorbing states $R_1,\ldots,R_n$, or the ``0'' state.
Observe that the event $\{T \leq t\}$ happens if and only if at time $t$ the process $x(t)$ is in one of 
the states $R_1,\ldots,R_n$, or in the ``0'' state. Therefore
\[
\Phi_T(t) \doteq \Prob\{T \leq t\} = \Prob\{ x(t) = R_1 \mbox{ or } x(t) = R_2 
\mbox{ or } \cdots \mbox{ or }  x(t) = R_n  \mbox{ or } x(t) = \mbox{``0''}
 \}.
\]
Since the events $\{x(t) = R_i\}$ are mutually exclusive, it holds that
\[
\Phi_T(t)  =  \Prob\{ x(t) = \mbox{``0''}\} + \sum_{i=1}^{n}\Prob\{ x(t) = R_i\}
=\pi_0(t) +  \sum_{i=1}^{n} \pi_i(t) = 
c_{n+1} \Upsilon_n(t) +
\sum_{i=1}^{n} c_i \Upsilon_i(t).
\label{eq:exittime_CDF}
\]
Since $T$ is nonnegative, the expected exit time can then be obtained as
\beas
\bar T =\bar T(\sigma) \doteq \E{T} &=&  \int_{0}^\infty (1-\Phi_T(t))\d t \\
&=&
=
 c_{n+1}\int_{0}^\infty  (1- \Upsilon_n(t))\d t +
 \sum_{i=1}^{n} c_i  \int_{0}^\infty  (1-\Upsilon_i(t))\d t \\
 &=& 
c_{n+1}\E{s_n} +    \sum_{i=1}^{n} c_i  \E{s_i}, 
\eeas
where, due to linearity of the expectation,
\beas
\E{s_i} &=& \E{\tau_{\sigma(1)} + \cdots + \tau_{\sigma(i)}} = 
 \E{\tau_{\sigma(1)}} + \cdots + \E{\tau_{\sigma(i)}} =  \theta_{\sigma(1)} + \cdots + \theta_{\sigma(i)}. 
 \eeas
Therefore,
\beq
\bar T(\sigma) =   c_{n+1}\sum_{k=1}^n \theta_{\sigma(k)} + \sum_{i=1}^{n} c_i \left( \sum_{k=1}^i \theta_{\sigma(k)}\right) .
\label{eq:barT}
\eeq
\section{Optimal opportunity scheduling}
For any given  scheduling sequence $\sigma$, we are now able to evaluate the eventual expected reward $\bar R$, given by eq.\ (\ref{eq:ssexpreward}), and the expected time at which the reward is obtained, $\bar T$, given by eq.\ (\ref{eq:barT}).
The problem we face next is to actually determine an optimal scheduling sequence $\sigma^*$ so to achieve an optimal tradeoff
between the reward value $\bar R$, which we like to be large, and the time value $\bar T$, which we like to be small, the ideal situation being to receive a large reward very soon in time.
The pair of values $(\bar T, \bar R)$ depends on the scheduling sequence $\sigma$. 
In principle, the player may compute
$(\bar T, \bar R)$ for all the $n!$ permutations of $\{1,2,\ldots,n\}$, and then pick the pair that best suits her.
However, this is only feasible for very small values of $n$. 

We quantify the tradeoff between the requests of
large $\bar R$ and small $\bar T$ by considering
 a composite objective function to be maximized.
Namely, we consider the problem
\beq
J^*(\eta) = \max_{\sigma\in \mbox{perm}\{1,\ldots,n\}} \; J(\sigma) \doteq \bar R(\sigma) -\eta \bar T(\sigma),
\label{eq:sched_opt}
\eeq
where $\eta\geq 0$ is some given tradeoff parameter, and the optimization is to be performed over
all permutations $\sigma$ of $\{1,\ldots,n\}$. By sweeping $\eta$ from the zero value to some large value, we obtain
all the tradeoff-optimal solutions. Geometrically, a tradeoff-optimal solution is a scheduling
$\sigma^*(\eta)$
that yields a point in the $(\bar T, \bar R)$ plane that passes through the line $\bar R -\eta \bar T = J^*(\eta)$,
where $J^*(\eta)$ is the optimal value of the objective in (\ref{eq:sched_opt}).

\begin{figure}[h!!]
\centerline{\scalebox{.68}{\includegraphics{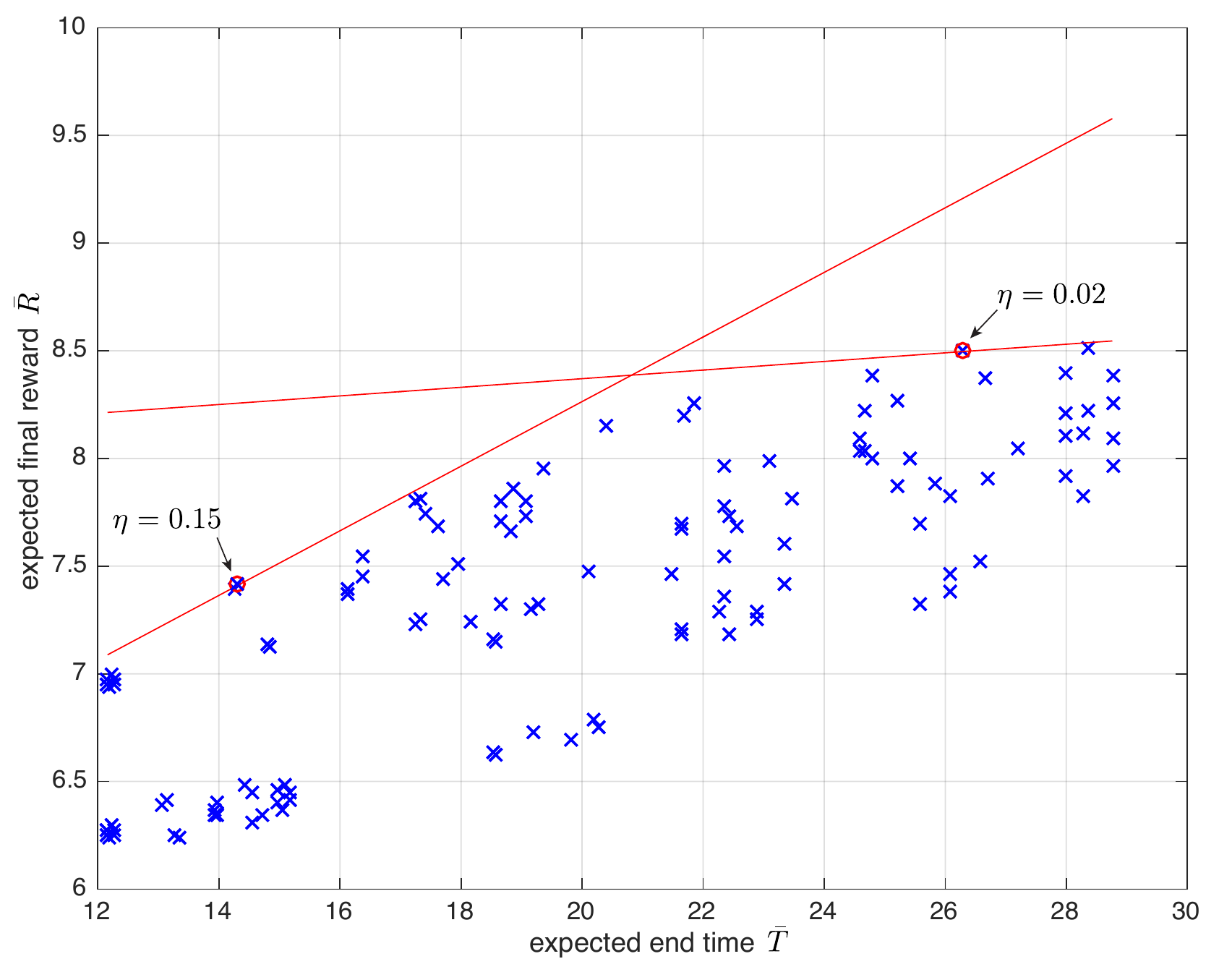}}}
\caption{Scatter plot of the pairs $(\bar T, \bar R)$ for the data in Table~\ref{tab_countex1}, for all the $n! = 120$ scheduling permutations.  }
\label{fig:scatter1}
\end{figure}

For example, using the data in Table~\ref{tab_countex1}, with
$n=5$, we evaluated $(\bar T, \bar R)$ for all the $n! = 120$ scheduling permutations, obtaining the scatter plot shown in Figure~\ref{fig:scatter1}, and two different optimal solutions corresponding to $\eta= 0.02$ and $\eta=0.15$.

\begin{table}[htb]
\begin{center}
    \begin{tabular}{ l||l | l | l }
$i$ & reward $r_i$ & probability $p_i$ & avg. time $\theta_i$  \\ \hline 
1 & 12    & 0.2    & 8     \\
2 & 10 & 0.3 & 14 \\
3 & 8.2    & 0.25 & 10  \\
4 & 8   &  0.5 & 5  \\
5 & 6    & 0.7 & 7 
\\
    \end{tabular}
    \end{center}
        \caption{\label{tab_countex1} Example  data.}
\end{table}

The tradeoff-optimal point in the upper-right part of the plot in Figure~\ref{fig:scatter1} is obtained from
 the value of the tradeoff parameter $\eta=0.02$, and corresponds the optimal scheduling $\sigma^*(\eta)=\{1,2,4,3,5\}$.
 The  point in the lower-left part of the plot is obtained from
the value of the tradeoff parameter $\eta=0.15$, and corresponds the optimal scheduling $\sigma^*(\eta)=\{4,1,5,2,3\}$.

Problem~(\ref{eq:sched_opt}) involves maximization of a nonlinear objective over 
permutations of the first $n$ integers and, as such, looks intractable as soon as $n$ grows beyond, say, the first ten integers
 (e.g., a brute-force approach for $n=10$ would already require checking $n! = 3,628,800$ scheduling  permutations).
Contrary to this first assessment, however, the main result below shows that optimal solutions
to problem~(\ref{eq:sched_opt}) can be very easily computed, by simply sorting a sequence of $n$ numbers. 
 
\begin{theorem}
\label{thm:main}
Any permutation $\sigma^*$ of $\{1,\ldots,n\}$ that is optimal for problem~(\ref{eq:sched_opt}) is such that
the sequence of numbers $\{r_{\sigma^*(i)} - \eta \theta_{\sigma^*(i)}/p_{\sigma^*(i)} \}$,
$i=1,\ldots,n$, is ordered non-increasingly.
\end{theorem}

\noindent
{\bf Proof.}
Consider the expression for the composite objective function
\[
J= \bar R -\eta \bar T = 
- \eta \prod_{j=1}^{n}(1-p_{\sigma(j)}) \sum_{k=1}^n \theta_{\sigma(k)} +
\sum_{i=1}^n z_i p_{\sigma(i)}\prod_{j=1}^{i-1}(1-p_{\sigma(j)}) ,
\]
where we defined
\[
z_i \doteq r_{\sigma(i)} -\eta  \sum_{k=1}^i \theta_{\sigma(k)},\quad i=1,\ldots,n.
\]
Observe that $z_i$ depends only on the value $r_{\sigma(i)}$, and on the
sum of the values of $\theta_{\sigma(1)},\ldots,\theta_{\sigma(i)}$;  hence, it is invariant
to permutations of these elements.  Further, observe that
the first term in the expression of $J$ does not depend on the scheduling sequence, hence it can be dropped from the 
objective without affecting the optimal solution. We thus consider only the last term in the objective, defining 
\beq
V_0(\sigma(1),\ldots,\sigma(n)) \doteq \sum_{i=1}^n z_i p_{\sigma(i)}\prod_{j=1}^{i-1}(1-p_{\sigma(j)}) 
\label{eq:V0}
\eeq
as the modified objective to be maximized.
Noticing that
\beas
\lefteqn{
V_0(\sigma(1),\ldots,\sigma(n))  =     z_{1} p_{\sigma(1)} + } \\
&& 
+(1-p_{\sigma(1)})\left(  
z_2 p_{\sigma(2)} +
 (1-p_{\sigma(2)}) \left(  
 z_3 p_{\sigma(3)}
 + (1-p_{\sigma(3)}) \left(  
 z_4 p_{\sigma(4)} + \cdots
 \right)
 \right)\cdots
  \right),
\eeas
we define
\beas
V_{k}(\sigma(k+1),\ldots,\sigma(n))  &\doteq & \sum_{i={k+1}}^n z_i p_{\sigma(i)}\prod_{j=k+1}^{i-1}(1-p_{\sigma(j)}),
\quad \mbox{for } k=0,1,\ldots,n-1,
\eeas
and we have that
\beas
V_0(\sigma(1),\ldots,\sigma(n))  &=& z_1  p_{\sigma(1)} + (1-p_{\sigma(1)}) 
V_{1}(\sigma(2),\ldots,\sigma(n)) \\
V_{1}(\sigma(2),\ldots,\sigma(n)) &=& z_2 p_{\sigma(2)} +
 (1-p_{\sigma(2)}) V_{2}(\sigma(3),\ldots,\sigma(n))  \\
 \vdots && \vdots \\
 V_{n-2}(\sigma(n-1),\sigma(n)) &=& z_{n-1} p_{\sigma(n-1)} +
 (1-p_{\sigma(n-1)}) V_{n-1}(\sigma(n))  \\
 V_{n-1}(\sigma(n))  &=& z_n p_{\sigma(n)}.
\eeas
Now, the problem of maximizing $V_0$
with respect to the scheduling sequence has ``optimal substructure,'' that is, it satisfies Bellman's Principle of Optimality
(see \cite{Bellman:57}, Sec.\ III.3), since
any optimal decision policy is such that
no matter what the initial $k$ decisions are, the remaining $n-k$
decisions constitute an optimal policy with respect to the state resulting from the initial decisions.
In our case, it is readily seen that if $\sigma^*(1),\sigma^*(2),\ldots,\sigma^*(n)$ is the  optimal sequence that maximizes $V_0$, then
$\sigma^*(2),\sigma^*(3),\ldots,\sigma^*(n)$  maximizes 
$V_{1}$, $\sigma^*(3),\ldots,\sigma^*(n)$ maximizes $V_{2}$, and so on.

Let then $\sigma^*$ be an optimal scheduling sequence, and consider, for any $k\geq 1$,
the optimal residual value
\beas
V_{k-1}^* &=& z^*_k p_{\sigma^*(k)} + (1-p_{\sigma^*(k)}) V_{k}^* \\
&=&
z^*_kp_{\sigma^*(k)} + (1-p_{\sigma^*(k)})
\left[ z^*_{k+1}p_{\sigma^*(k+1)}  + (1-p_{\sigma^*(k+1)}) V_{k+1}^* \right].
\eeas
We next evaluate the residual value $V_{k-1}'$ obtained by exchanging 
the order of the $k$-th and the $(k+1)$-th decisions in the optimal scheduling sequence.
Exchanging $\sigma^*(k) \leftrightarrow \sigma^*(k+1)$ yields
\beas
z'_k &\doteq &  z^*_{k+1} +\eta \theta_{\sigma^*(k)}\\
z'_{k+1} &\doteq  &  z^*_{k} -\eta \theta_{\sigma^*(k+1)},
\eeas
whence
\beas
V_{k-1}' &=& z'_k  p_{\sigma^*(k+1)} + (1-p_{\sigma^*(k+1)})
\left[z'_{k+1} p_{\sigma^*(k)}  + (1-p_{\sigma^*(k)}) V_{k+1}' \right] \\
&=& z'_k  p_{\sigma^*(k+1)} + (1-p_{\sigma^*(k+1)})
\left[z'_{k+1} p_{\sigma^*(k)}  + (1-p_{\sigma^*(k)}) V_{k+1}^* \right] ,
\eeas
where we used the fact that $V_{k+1}'  = V_{k+1}^*$, since the calling sequence
from $k+2$ onwards is kept unchanged, and $V_{k+1}$ is invariant 
to permutations in the ordering of the previous scheduling, from $1$ up to $k+1$.
Straightforward computations then show that
\beas
  V_{k-1}^*  - V_{k-1}'  &=& 
z^*_kp_{\sigma^*(k)} + (1-p_{\sigma^*(k)})
\left[ z^*_{k+1}p_{\sigma^*(k+1)}  + (1-p_{\sigma^*(k+1)}) V_{k+1}^* \right] \\
&& - z'_k  p_{\sigma^*(k+1)} - (1-p_{\sigma^*(k+1)})
\left[z'_{k+1} p_{\sigma^*(k)}  + (1-p_{\sigma^*(k)}) V_{k+1}^* \right]  \\
&=&
z^*_kp_{\sigma^*(k)} +  z^*_{k+1}p_{\sigma^*(k+1)}   -p_{\sigma^*(k)}z^*_{k+1}p_{\sigma^*(k+1)} +
 (1-p_{\sigma^*(k)}) (1-p_{\sigma^*(k+1)}) V_{k+1}^*  \\
 && 
 - z'_k  p_{\sigma^*(k+1)} - z'_{k+1} p_{\sigma^*(k)} + p_{\sigma^*(k+1)}z'_{k+1} p_{\sigma^*(k)}
 - (1-p_{\sigma^*(k+1)}) (1-p_{\sigma^*(k)}) V_{k+1}^*  \\
 &=& 
 p_{\sigma^*(k)}(z^*_k- z'_{k+1}) + p_{\sigma^*(k+1)} ( z^*_{k+1}- z'_k) + 
 p_{\sigma^*(k)}p_{\sigma^*(k+1)} (z'_{k+1} -z^*_{k+1} ) \\
 &=& 
 p_{\sigma^*(k)}\eta \theta_{\sigma^*(k+1)} 
 - p_{\sigma^*(k+1)} \eta \theta_{\sigma^*(k)} 
 + p_{\sigma^*(k)}p_{\sigma^*(k+1)} (r_{\sigma^*(k)}-r_{\sigma^*(k+1)} ) \\
 &=& 
  p_{\sigma^*(k)}p_{\sigma^*(k+1)}
   \left( 
   (r_{\sigma^*(k)} - \eta \theta_{\sigma^*(k)}/p_{\sigma^*(k)})
   - (r_{\sigma^*(k+1)}  -\eta  \theta_{\sigma^*(k+1)}/p_{\sigma^*(k+1)}    )
   \right).
\eeas
Now, since $V_{k-1}^* $ is maximal by definition, it must hold that
$V_{k-1}^*  - V_{k-1}' \geq 0$, thus
\[
 p_{\sigma^*(k)}p_{\sigma^*(k+1)}
   \left( 
   (r_{\sigma^*(k)} - \eta \theta_{\sigma^*(k)}/p_{\sigma^*(k)})
   - (r_{\sigma^*(k+1)}  -\eta  \theta_{\sigma^*(k+1)}/p_{\sigma^*(k+1)}    )
   \right) \geq 0,
\]
which  implies that 
\[
r_{\sigma^*(k)} - \eta \theta_{\sigma^*(k)}/p_{\sigma^*(k)} \geq 
r_{\sigma^*(k+1)}  -\eta  \theta_{\sigma^*(k+1)}/p_{\sigma^*(k+1)},
\]
i.e., that any optimal scheduling sequence is such that the values 
in the sequence $\{r_{\sigma^*(i)} - \eta \theta_{\sigma^*(i)}/p_{\sigma^*(i)} \}$,
$i=1,\ldots,n$,
are ordered non-increasingly.
\qed

\begin{remark}[Tradeoff-optimal and Pareto-optimal points] \rm
\label{rem:frontier}
Consider all possible perm\-uta\-tions of $\{1,\ldots,n\}$ and plot the corresponding 
$\bar R$ and $\bar T$ values as points in the
$(\bar T,\bar R)$ plane, as we did in in Figure~\ref{fig:scatter1}.
The convex hull of such points is a polytope. The tradeoff-optimal points that we find by maximizing 
the objective $J=\bar R -\eta \bar T$ correspond to vertices of this polytope. More precisely, they correspond to the
 farthest points in the direction $[-\eta,\; 1]$ that are tangent to the polytope. All such vertex solutions can be found via Theorem~\ref{thm:main}, by sweeping over a suitable grid of values of the tradeoff parameter $\eta$.
 We remark, however, that there may exist points that are optimal in a Pareto sense, but that cannot be found via the proposed method. Pareto-optimal points in the $(\bar T,\bar R)$ plane are all those points for which no other point exists
 that is simultaneously to the West and to the North of the considered point. For such points it is indeed not possible to simultaneously increase the expected reward {\em and} decrease the expected end time. Clearly, all the vertex points
 that we find by maximizing $J$ are Pareto-optimal, but there may exist Pareto-optimal points that lie in the interior of the polytope and that hence cannot be found via Theorem~\ref{thm:main}. Some of these points can easily be spotted by observing Figure~\ref{fig:scatter1}.
\end{remark}

\begin{remark}[Batch vs.\ sequential schedule] \rm
\label{rem:sequential}
In a typical dynamic decision problem one may take two approaches. The first one is a ``batch'' approach, in which
all decisions are taken  at time $\tau_0=0$, by maximizing the objective and thus computing a-priori all decisions
to be implemented at later times.
However, at time $\tau_0 = 0$ we really need to know and implement only the first decision (i.e., in out context, find the first opportunity to try). For the second decision, we can {\em wait and see} what is the outcome of
the first opportunity, and only then decide which is the one to try second, and so on for later decisions.
Such as sequential approach  is, in general, better than the batch one, since the player can exploit 
full information from the realizations of the uncertainty that happened right before the time when the decision is really needed.

The approach we took in this work is a batch one. However, it is easy to prove that, for the specific problem at hand, it makes no difference to take a batch or a sequential decision approach. Indeed, in a sequential approach, the first decision
$\sigma^*(1)$ is computed by maximizing the full objective $V_0$ in (\ref{eq:V0}), hence the first decision
is the same for the batch and the sequential approach.
The sequential approach would then consider the conditional value of the objective, given the outcomes of the first decision $\sigma^*(1)$.
It is straightforward to verify that the conditional value of the forward objective after the outcome of the first opportunity is 
of the form $V_1$, plus additive terms that depend on the first decision and that do not affect the optimal solution.
The optimal second decision, according to the sequential approach, is the one that maximizes $V_1$, and hence it coincides with the solution $\sigma^*(2)$ from the batch approach. Similarly, at any later decision time $\tau_k$, the sequential approach  maximizes a conditional objective which coincides with $V_k$ (besides additive terms that do not affect the optimal solution), hence the ensuing sequential decisions $\sigma^*(k)$ coincide with the batch decisions.
\end{remark}

\subsection{Example}
We considered a second example with $n=20$ opportunities,
with the randomly generated data shown in the data matrix below, in which the first row contains the rewards
$r_i$, the second row contains probabilities $p_i$, and the third row contains  mean response times $\theta_i$.

\begin{figure}[h!tb]
\centerline{\scalebox{.75}{\includegraphics{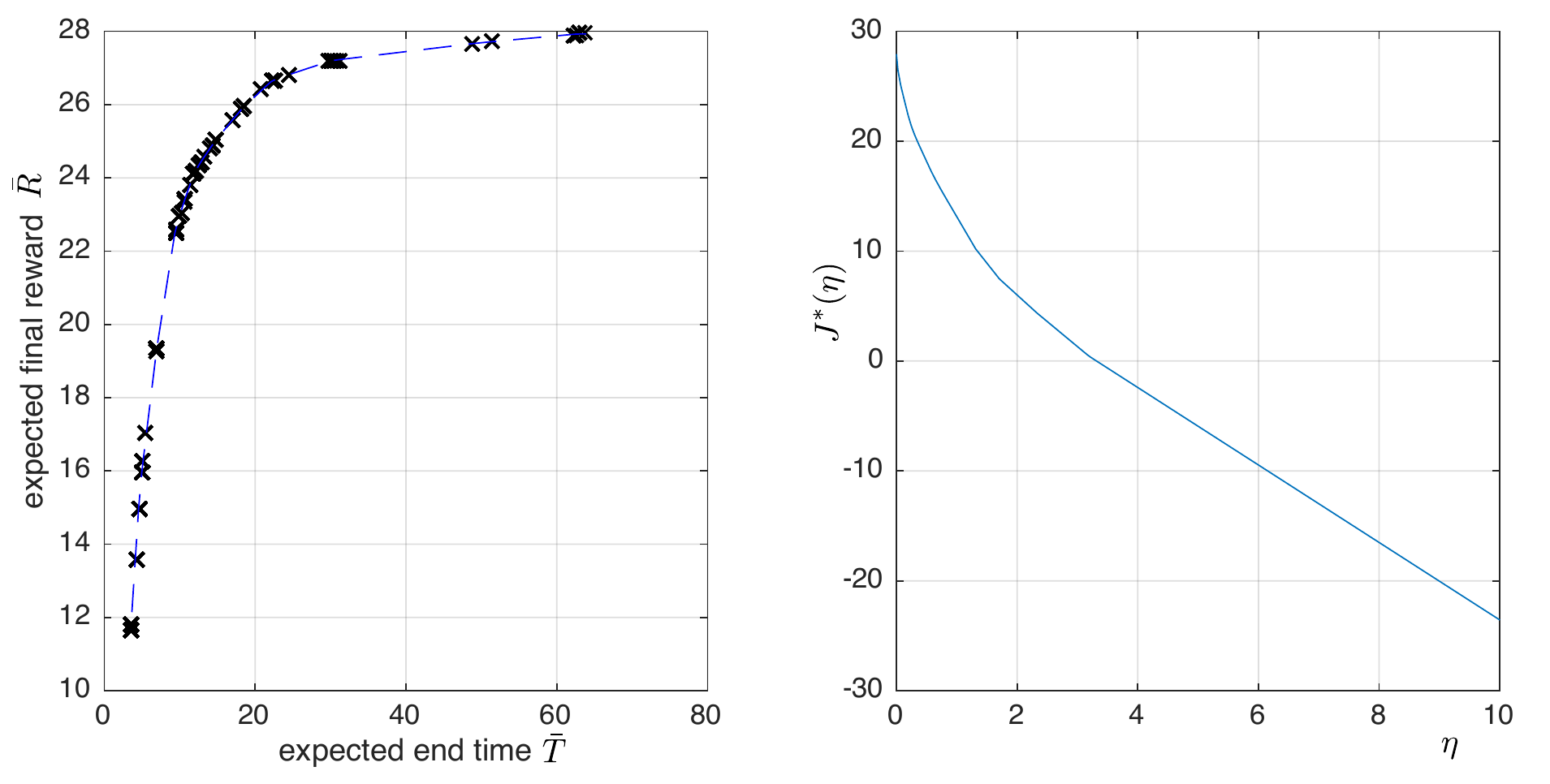}}}
\caption{Left: $(\bar T, \bar R)$ pairs on the tradeoff-optimal frontier, obtained for values of the tradeoff parameter
$\eta\in[0,\, 10]$. Right: plot of the optimal objective value $J^*(\eta)$.
 }
\label{fig:exampleplot}
\end{figure}

\beas
O &=&
\left[\begin{array}{ccccccccccc} 27.0 & 14.0 & 3.0 & 28.0 & 9.0 & 19.0 & 20.0 & 7.0 & 29.0 & 20.0 & 17.0  \\
0.449 & 0.659 & 0.753 & 0.805 & 0.029 & 0.780 & 0.567 & 0.076 & 0.252 & 0.133 & 0.564 \\
18.0 & 2.0 & 2.0 & 43.0 & 11.0 & 21.0 & 7.0 & 13.0 & 27.0 & 15.0 & 4.0
\end{array}\right. \\
&&
\left.\begin{array}{ccccccccc}
27.0 & 17.0 & 1.0 & 15.0 & 25.0 & 23.0 & 18.0 & 1.0 & 22.0\\ 
0.541 & 0.069 & 0.988 & 0.251 & 0.315  & 0.300 & 0.042 & 0.528 & 0.256\\ 
7.0 & 33.0 & 37.0 & 20.0 & 3.0 & 8.0 & 10.0 & 40.0 & 51.0 \end{array}\right].
\eeas
A brute-force approach that explores all permutations of $\{1,\ldots,20\}$ is in this case unviable, since there are
$20! \simeq 2.43 \times 10^{18}$ permutations.
However, by using Theorem~\ref{thm:main} we readily find all solutions on the tradeoff-optimal frontier, for
gridded values of $\eta$ in the interval $[0,\, 10]$, see Figure~\ref{fig:exampleplot}.

\begin{figure}[h!tb]
\centerline{\scalebox{.7}{\includegraphics{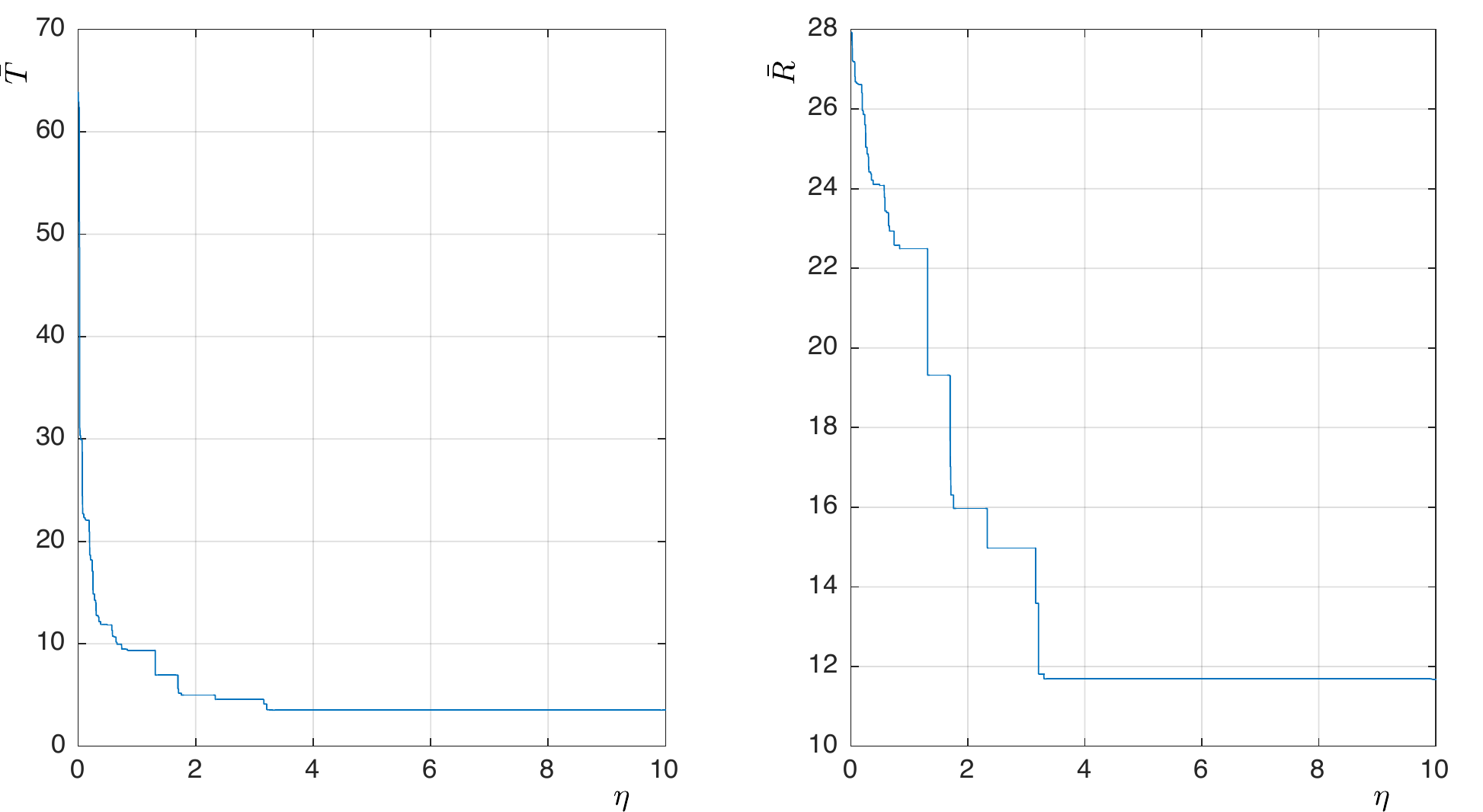}}}
\caption{Left:  value of $\bar T^*$ as a function of $\eta$.
Right: value of $\bar R^*$ as a function of $\eta$.
 }
\label{fig:RTplot}
\end{figure}

For $\eta=0$, we find the optimal value $J^*(0) = 27.928$,
with $\bar R^*(0) =27.928 $, $\bar T^*(0) = 63.91$, and
optimal sequence
\[
\sigma^*(0) = \{ 9,     4,    12,     1,    16,    17,    20,     7,    10,     6,    18,   11,    13,    15,     2,     5,     8,     3,    14,    19\},
\]
which simply corresponds to ordering the rewards $r_i$ non-increasingly, regardless of all other parameters.
Notice that since the time $\bar T$ does not influence the objective $J$, for $\eta=0$, the optimal scheduling sequence has a high expected end time $\bar T^*(0) = 63.91$.
However, for example, to a value $\eta =0.5 $, it corresponds a value of $\bar R^*(0.5) = 24.08$,
while $\bar T^*$ drops to  $\bar T^*(0.5) = 11.84$, that is we reduce the expected time by about $81\%$
at the price of a modest $13.8\%$ decrease in expected reward, see Figure~\ref{fig:RTplot}.
The optimal sequence, for  $\eta =0.5 $, is
\[
\sigma^*(0.5) = \{  12,    16,     7,    11,     2,    17,     1,     6,     3,     4,    14,     9,    15 ,   10 ,   19,    20,     8,    18,     5,    13\}.
\]

\section{Discussion}
\label{sec:discussion}
A quick analysis of the literature reveals that our Dating Problem has indeed connections to several more or less known problems.
A classical one, apparently known since the 1950s, is the so-called Best Choice Problem,
also known as 
the Secretary Problem which, incidentally, was about hiring, not dating, the secretary; see \cite{Ferguson:89}. 
The structure of this problem is, however, different from the Dating Problem
in that opportunities are presented to the decision maker in random order, one at a time, and she has to select the best of  all opportunities. Closer to our model is a problem described in \cite{Duffuaa:96}, which extends previous results from
\cite{Alidaee:94}, and which
 deals with multi-characteristic sequential testing and job processing on a single machine.
 In \cite{Duffuaa:96}, a stochastic ``Candidate Selection Problem'' is described, in which $n$ candidates are to be selected for filling one available position. The benefits for hiring candidate $i$ are quantified by a random variable $r_i$,
 and the probability of acceptance of the job by candidate $i$ is $p_i$. There is a cost for offering the job described by a random variable $C$, and the objective is to find the optimal selection schedule of the candidates in order to maximize the expected benefit. The main difference with respect to the Dating Problem is that the response times of the candidates are not taken into consideration in the model and as  components of the objective to be maximized. 
The management literature also has other interesting examples, structurally quite unrelated to the Dating Problem, 
but for which optimal decision are obtained by simply putting some sequence of numbers in monotonic order.
One example is the so-called Smallest Variance First rule for minimizing the expected makespan in a single-machine scheduling problem, see \cite{Pinedo:07}. Another example involves lot sizing in sequential English auctions, where
a widely held notion is that lots should be arranged in order of decreasing value, see, e.g., \cite{TrNaKa:09} and
\cite{Mitha:12}.

The basic Dating Problem presented here can be extended in several ways. One direction could be that of relaxing the assumption that the decision maker has exact knowledge of the parameters $(r_i,p_i,\theta_i)$, thus considering {\em ambiguity} in the stochastic description of the problem.


\end{document}